# Stable, intense 1 kHz supercontinuum light generation in air

Yaoxiang Liu,[1,2,3] Tie-Jun Wang,[1,2,*] Hao Guo,[1,2] Na Chen,[1] Xuan Zhang,[1] Haiyi Sun,[1] See Leang Chin,[4] Yuxin Leng,[1,2] Ruxin Li,[1,2,*] and Zhizhan Xu[1,2]

[1]*State Key Laboratory of High Field Laser Physics, Shanghai Institute of Optics and Fine Mechanics and CAS Center for Excellence in Ultra-intense Laser Science, Chinese Academy of Sciences, Shanghai 201800, China*
[2]*Center of Materials Science and Optoelectronics Engineering, University of Chinese Academy of Sciences, Beijing 100049, China*
[3]*MOE Key Laboratory of Advanced Micro-structured Materials, Institute of Precision Optical Engineering, School of Physics Science and Engineering, Tongji University, Shanghai 200092, China*
[4]*Centre d'Optique, Photonique et Laser (COPL) and Département de physique, de génie physique et d'optique, Université Laval, Québec, Québec G1V 0A6, Canada*
*tiejunwang@siom.ac.cn (TJW), and ruxinli@siom.ac.cn (RXL)

**Abstract:**

Supercontinuum (SC) light source has advanced ultrafast laser spectroscopy in condensed matter science, biology, physics, and chemistry. Compared to the frequently used photonic crystal fibers and bulk materials, femtosecond laser filamentation in gases is damage-immune for supercontinuum generation. A bottleneck problem is the strong jitters from filament induced self-heating at kHz repetition rate level. We demonstrate stable kHz supercontinuum generation directly in air with multiple mJ level pulse energy. This is achieved by applying an external DC electric field to the air plasma filament through the effects of plasma wave guiding and Coulomb interaction. Both pointing and intensity jitters of 1 kHz air filament induced SC light are reduced by more than 2 fold. This offers the opportunities for stable intense SC generation and other laser filament based applications in air.

**Introduction**

The supercontinuum (SC) light source comes from the generation of intense ultrafast broadband pulses from the nonlinear interaction and propagation of ultrafast laser pulses in an optical transparent medium. It is also called "white light" laser when the laser spectrum covers from the near ultraviolet to the infrared wavelengths. The invention and development of SC laser [1] has advanced many applications, for instance, in ultrafast laser pulse generation and compression (attosecond pulse) [2], extremely high-precision optical frequency and time metrology [3, 4], terabits/sec high-capacity information communication [5, 6] and ultrafast laser spectroscopy [7, 8] and imaging [9]. Towards the applications, the SC light sources are mostly generated from bulk materials [10] and photonic crystal fibers [11] which limits the SC pulse energy due to the low damage threshold of solid materials. Ultrafast laser filamentation in gases [12-17] is immune from the materials' damage for SC generation. However, the milliseconds thermal diffusion in an air filament [18-20] leads to air density depletion at the arrival of the next laser pulse for kHz repetition rates filamenting laser. The thermal self-action effect induces unstable filament beam pointing [21-24] which sets a challenge for high repetition rate filament based SC light source and its applications.

In this work, we report a stable, intense, high repetition rate SC generation through filamentation directly in air. This is accomplished by simply applying an external DC electric field on the filament plasma channel. Through the combined effects of plasma wave guiding [25, 26] and Coulomb interaction, both pointing and intensity jitters of 1 kHz air filament induced SC light are reduced by more than 2 fold. 3.55 mJ stable SC pulse at 1 kHz is achieved with a 6.54 mJ filamenting multi-mode pulse directly in air. This technique on reducing the thermally induced instability of high repetition rate filaments would work at other filamenting laser wavelengths due to the plasma nature and therefore provides new opportunities for intense stable SC generation and its applications. The stable air filament should directly benefit other filament-based applications, for instance, in remote sensing [8, 27], imaging [28] and micro-machining of condensed materials [29, 30], etc.

## Results

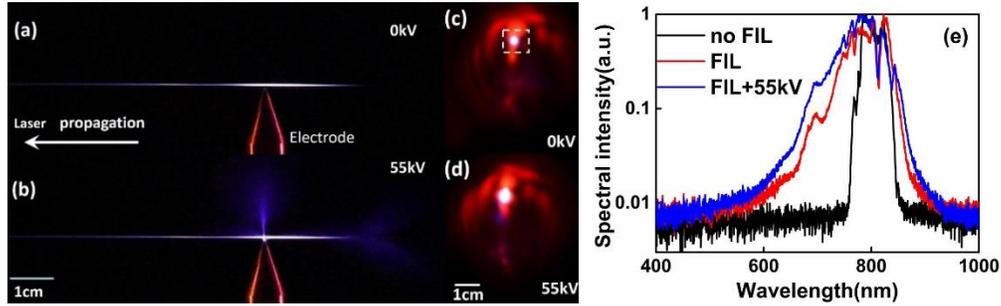

Fig. 1. Real color images of laser filament in air with high voltage off (a) and on (b). The corresponding forward beam patterns on a white screen in real color: (c) is with high voltage off and (d) is with high voltage on. Laser pulse energy was 6.12 mJ and the voltage applied on the electrode was 55 kV. The laser filament and the electrode are in the horizontal plane. (a) and (b), (c) and (d) have the same length scale, respectively. (e) the typical white light spectra after filamentation with (FIL+55kV) and without (FIL) external DC electric field together with the initial laser spectrum (no FIL) for comparison. Each spectral distribution was normalized at its maximum.

In the experiments, by focusing 34fs/6.12mJ laser pulses at 1 kHz in air with an f=50 cm plano-convex lens, a 9 cm long filament (Fig. 1(a)) or a white light laser (the white spot inside the dashed square in Fig. 1(c)) was generated. The electrode was placed perpendicularly to the filament in the horizontal plane. When applying a DC voltage of 55 kV on the electrode, laser guided coronas [31, 32] were observed (Fig. 1(b)). Far field (approximately 1.2 m to the filament) beam pattern recorded by the digital camera is shown in Fig. 1(d). The spectra of white light laser after filamentation with (FIL+55kV) and without (FIL) external DC electric field are shown in Fig. 1(e) together with the initial laser spectrum (no FIL) for comparison. The spectra of the FIL and FIL+55kV were significantly broadened as compared with the pump laser pulse which is due to self-phase modulation (SPM) [1, 15]. The SC bandwidth was further broadened when the external DC high voltage was applied on the electrode. The spatial positions of the white light laser recorded frame by frame (60 fps) by the camera are shown in Fig. 2. It is obvious that the pointing of white light laser from the filament (DC electric field was off) appeared in a larger area (Fig. 2(a)) due to the wandering nature of the high repetition rate filament [21-24].The voltage was tuned from 0 kV up to 55 kV in the experiment; i.e. the electric field around the electrode tip changed from 0 V/m to $6.6\times10^7$ V/m; the electric field was obtained by solving the Poisson equation and Maxwell equation through finite element analysis. Once the DC electric field was applied near the filament, the white light laser pointing scattering became smaller (Fig. 2(b)-2(f)).

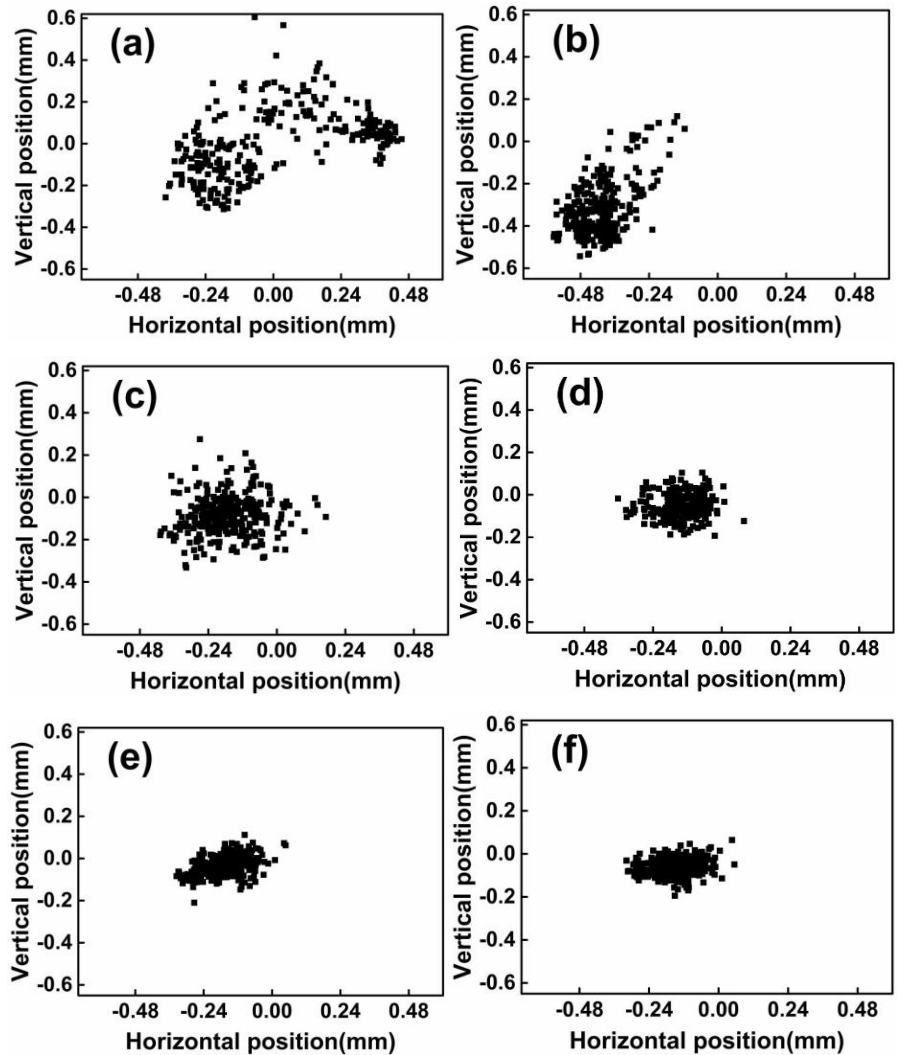

Fig. 2. Maps of forward 300 white light spots under different external DC electric field around the electrode: (a) 0 V/m, (b) $2.4\times10^6$ V/m, (c) $3\times10^7$ V/m, (d) $4.2\times10^7$ V/m (e) $5.4\times10^7$ V/m, (f) $6.6\times10^7$ V/m. The statistic average center point of the white light without voltage is defined as (0, 0).

We defined the statistical center point of the 300 shots of white lights from the filaments at 0 V/m electric field as (0, 0). The reference axis was therefore the line between this center point and the intersection point between the electrode's pointing direction and the filament. The standard deviations (SDEV) of the SC laser scattering (Fig. 2) with respect to the reference axis were calculated to depict the pointing stability. The SDEVs were normalized to the one at zero electric field when the electric field is applied. Compared with the white light pointing stability from laser filament only (external DC electric field of 0 V/m), the pointing stability in the horizontal plane was improved by approximately 3 times and tended to be stable once the external DC electric field was applied (Fig. 3(a)). Similar improvement was observed when negative high voltage was applied to the electrode. Moreover, similar results were obtained when the direction of laser's linear polarization was changed from the horizontal to the vertical, i.e. the pointing stability improvement of the white laser by the electric field was impregnable to the laser polarization direction.

Besides, the spatial pointing position of the white light laser can be tuned by the external DC electric field. In the case of positive voltage on the electrode, once the DC field was turned on, the center of gravity of the white light laser beam scattering 'suddenly' moved significantly away from the zero field position (Fig. 2b) through coulomb interaction between the electric field and the filament. Afterwards, when the DC field was further increased, the displacement decreased and was proportional to the voltage (effective electric force). The displacement stabilized (Fig. 2 c-f) when the threshold for corona generation ($\sim 2.6 \times 10^6$ V/m) [33] was reached. The detailed displacement is shown in Fig. 3b (points). Numerical simulation based on the Coulomb interaction is in good agreement with the experimental results (Fig. 3(b), curve).

Since the spatial jitter of laser filaments was principally due to the self-induced thermal heating, the higher the filament repetition rate was, the stronger the spatial jitter would be (the cases of "0" voltage in Fig. 3(c)). The DC electric field control technique could significantly improve the pointing stability by reducing the spatial jitters from filament self-heating, which was especially good for higher repetition rate, for example, at 1 kHz (Fig. 3(c)). Furthermore, the SDEV of the integrated SC spectral intensities (from 650nm to 900nm) evidenced the significant improvement (more than 3 fold) of SC intensity stability (Fig. 3(d)). As the electric field increased, the SDEV decreased, in which negative DC field worked better due to the difference in the coupling between the laser filament and the DC electric field. Compared with the positive DC electric field, the filament guided negative corona was confined along the plasma channel and was more stable [32]. By using our approach, 3.55 mJ stable SC pulses at 1 kHz was obtained when 6.54 mJ input pulse underwent filamentation directly in air with a 1m external focusing lens (Fig. 3(e)).

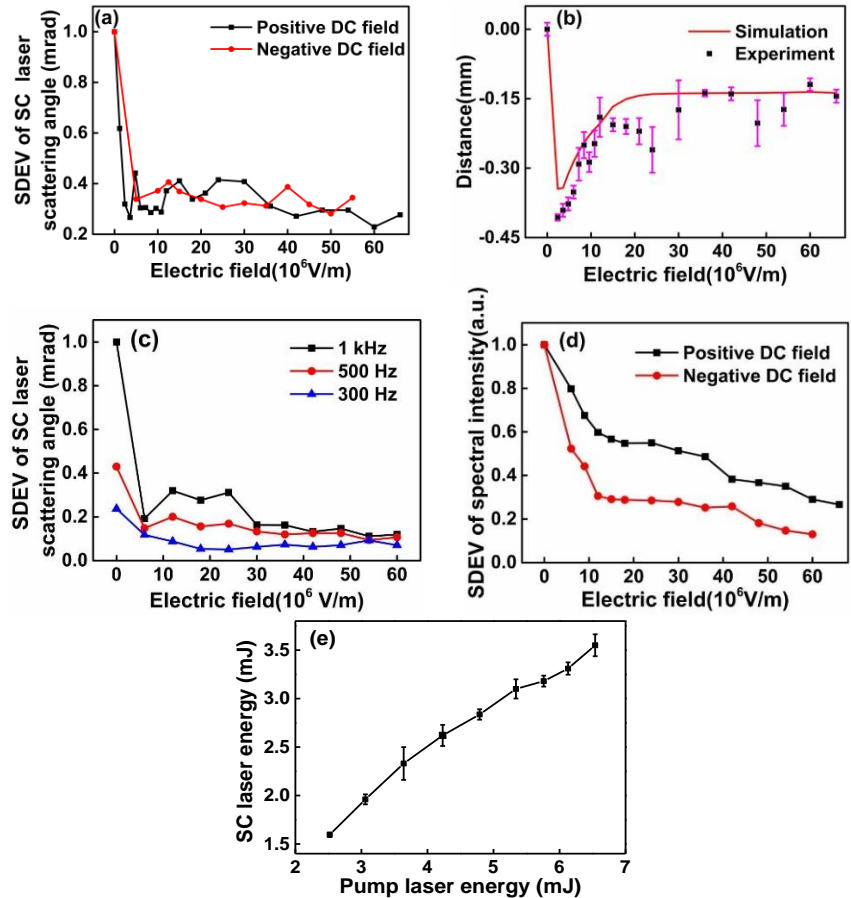

Fig. 3 (a) SDEV of the SC light forward pointing angle at 1 kHz as a function of the external DC electric field. (b) Experimental and numerical results of white light displacement distance in horizontal direction under different positive electric fields from initial position (0 V/m). Position '0' represents the initial mean position without the DC electric field and negative displacement is defined as when the filament is closer to the electrode. (c) The dependence of the SDEV of the SC light forward pointing angle on laser repetition rate in the horizontal direction. (d) SDEV of the integral SC spectral intensities working at 1 kHz under different polarities (positive and negative) DC electric fields. The integral spectral range is from 650 nm to 900 nm. The SDEVs are normalized to the one at zero electric field. (e) The SC laser energy obtained as a function of pump laser energy under 1m focusing condition.

Since the SC laser was generated through filamentation in air, the stable SC laser should originate from the localization of air filament under the external DC electric field. By inserting a flat ceramic plate in the high intensity plasma channel, the area of filament ablated patterns on the ceramic plate (Fig. 4(a) & 4(b)) was another indicator of filament pointing stability. By accumulating 7500 shots (7.4mJ/34fs/1kHz) on the plate, filament ablated areas as a function of the voltage is shown in Fig. 4(c). The smaller area under the external DC electric field clearly conformed with the improvement of filament pointing stability in air. Moreover, the filament ablated area varied with the laser repetition rate (the cases of 0 V/m electric field in Fig. 4(c)). The jitter of filament pointing was bigger at higher repetition rate due to filament induced self-heating [21-24]. Adding the DC electric filed clearly stabilized filament pointing, especially, at higher laser repetition rate. This DC electric field assisted stabilization had much less effect at repetition rate below 100 Hz (Fig. 4(c)).

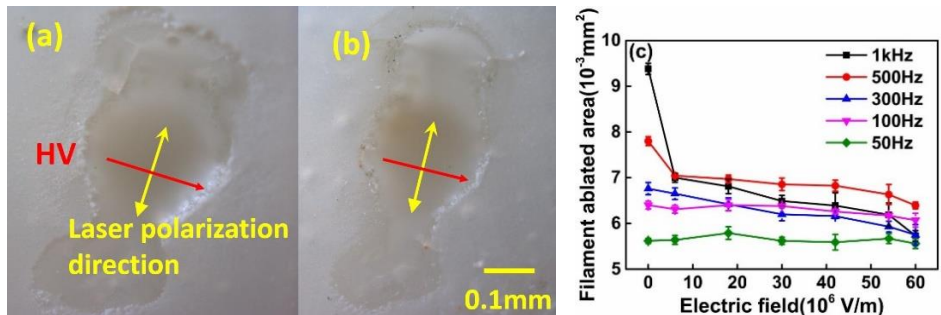

Fig.4 1 kHz laser filament ablated patterns on the ceramic samples after 7500 shots: (a) 0 kV, and (b) 5 kV ($6\times10^6$ V/m). Laser energy was set at 7.5 mJ and the frequency-was fixed at1 kHz. (c) shows the filament ablated areas under different repetition rates.

During femtosecond laser filamentation in air, initially there is a plasma density maximum on the optical axis due to laser instantaneous ionization by the initial Gaussian distribution of laser intensity. Typically, the lifetime of the plasma is within few nanoseconds time scale in an air filament. The thermal self-heating of laser filament comes mainly from the high field ionization and electron-ion recombination. The long thermal decay time (ms) becomes non-negligible at high laser repetition rate. We have carried out a simulation of the interaction. When an external DC electric field is applied on the laser filament, the effect of the external large DC electric field is three fold. Firstly, the length of the plasma filament is extended with slightly higher laser intensity (Fig. 5(a)) and higher plasma density (Fig. 5(b)), which favors broader SC generation through SPM (Fig. 1(e)). Secondly, the lifetime of the plasma is also significantly extended (Fig. 5(c)). The extended lifetime of the plasma is long enough (around 1 ms) (Fig. 5(d)) to bridge the interval time between two successive pulses. Thus, external DC electric field assisted plasma self-guiding effect contributes to the stability improvement of laser filament as well as the filament based white light laser in air. Finally, due to the plasma property of laser filament, external DC electric field provides an extra-control on the spatial localization of laser filament through Coulomb interaction (Fig. 3(b)).

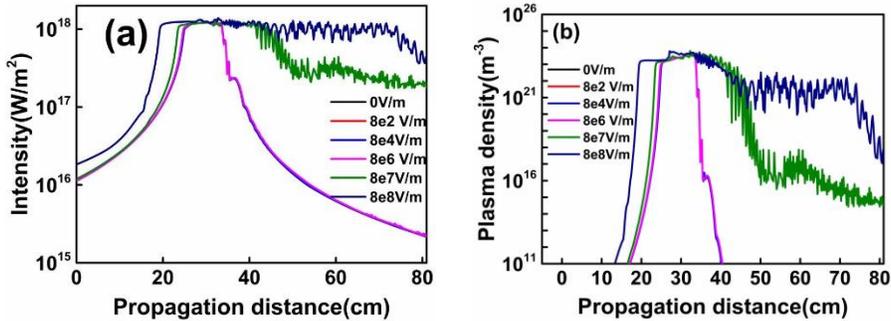

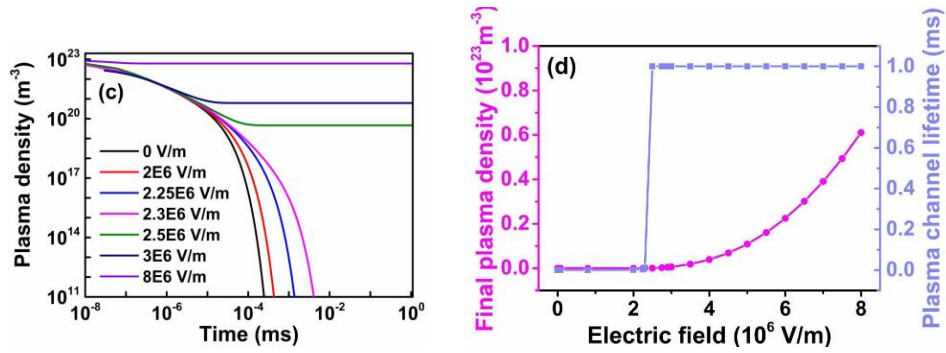

Fig. 5 (a), (b) and (c) are simulated on-axis laser intensity, plasma density and its decay time, respectively, under different electric field strengths according to the experimental parameters. (d) is the simulated plasma lifetime and density as a function of external DC electric field strength.

## Discussion

From the physical point of view, many electrons inside the filament are attracted towards the electrode in the case of positive HV, for example, while the positive ions are pushed away from the filament zone. This would significantly reduce the recombination rate inside the filament; hence, much less thermal jitter of the beam pointing direction. In the case of negative HV, many electrons are pushed away from the electrode while the positive ions move towards the electrode. Again, the recombination rate inside the filament is reduced; hence, thermal jitter of the beam pointing direction is reduced. Moreover, since the improvement of the stability of the white light laser by the electric field is mainly due to the decrease of the recombination between electrons and ions in the electric filed, laser polarization state would not affect the charges' spatial distribution. Thus, the improvement of pointing stability of the filament is independent of the laser polarization. Furthermore, the displacement of the filament as a whole with respect to the electrode can be explained as follows. Once a small DC field, positive for example, is applied, some electrons inside the filament will be pulled towards the electrode while the positive charges are repelled. The Coulomb attraction between the positive charges and the electrons will keep the filament intact. The filament as a whole will thus be pulled towards the electrode and the filament diameter becomes larger (compare Fig. 1 (d) and (e)). When the voltage is increased, some electrons will be pulled out of the filament into the electrode. Consequently, the extra positive charges in the filament will be pushed out of the filament. The plasma density inside the filament will now be lower; hence, the attractive force exerted by the electrode is smaller. The filament displacement is thus smaller. Consequently, the higher the voltage (electric field) is, the smaller the displacement will be (Fig. 3(b)). When corona discharge (avalanche ionization) starts, the strong plasma at the tip of the electrode will partially shield the field from attracting more electrons into the electrode. Thus, the electric attractive force stays constant and the displacement of the filament becomes constant (Fig. 3b).

In summary, we have experimentally demonstrated stable SC light source generation through femtosecond laser filamentation in air. By applying a large external DC electric field on the plasma filament in air, both pointing and intensity jitters of 1 kHz SC light are reduced by more than 2 fold. The stable SC pulse with energy of 3.55 mJ at 1 kHz is achieved by using 6.54 mJ filamenting pulse directly in air. Our approach solves the current standing problem of high repetition rate filament associated strong jitters, which works at other filamenting wavelengths due to the plasma nature. The findings presented here not only paves a way to generate intense high repetition rate SC light through damage-immune gases and should be beneficial for SC application, but also are crucial and useful for other filament-based applications on remote sensing and remote air lasing, atmospheric analysis, imaging and micromachining of condensed materials etc..

## Methods

### Experimental arrangement

A Ti:sapphire laser pulse (central wavelength at 800 nm) was focused by a f=50 cm lens to create filaments in air. DC high-voltage generator produced the voltage (-/+ max. 100 kV/ max. 1000 W) was connected to a copper cylindrical electrode with a sharp tip (the radius of curvature of the electrode tip was 0.68 mm. The axis of the electrode was perpendicular to the laser propagation direction in the same horizontal plane with a distance of approximately 1 mm between the tip and the laser filament. The laser polarization direction is in the vertical direction. The experiments were conducted inside a homemade Faraday cage ($0.5 \times 0.5 \times 0.4$ m$^3$) which is safely grounded. The forward white light laser generated by the filament in air was scattered by a white diffusing plate at the far field (1.2 m from the filament). A focusing lens (2-inches in diameter, f= 60 mm) set at an angle of 30° to the laser propagation axis imaged the scattered light to a fiber coupled spectrometer (Ocean Optics HR4000) for spectral analysis. A neutral density filter was put in front of the spectrometer to attenuate the diffused intensity. A digital camera (Nikon D7200) with 60 fps was used to monitor the spatial stability of the white light laser on the plate.

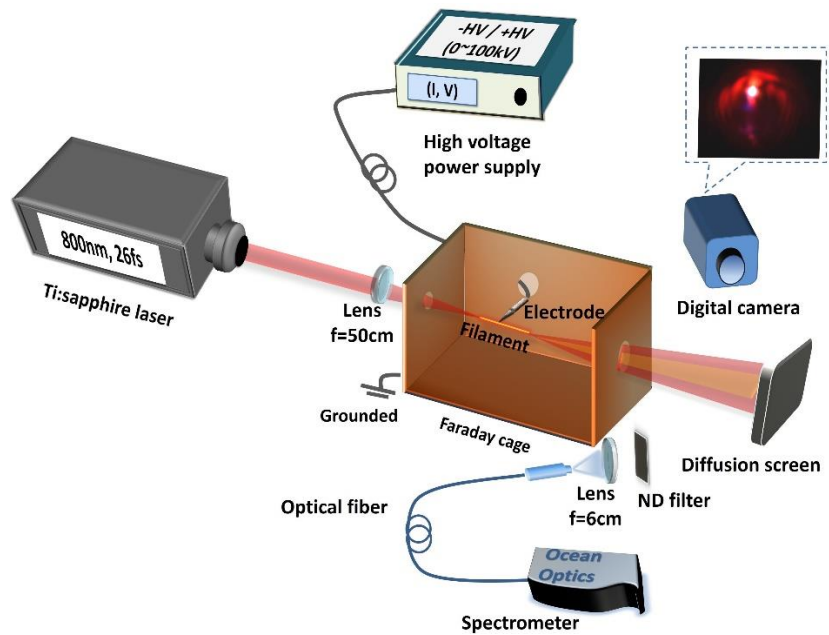

Fig. s1. Schematic of experimental setup.

## Simulation on DC electric field biased filamentation

The nonlinear Schrödinger equation (NLSE) is used to simulate the filamentation process. Input laser pulse is assumed to be an axially symmetrical Gaussian beam. When applying an external DC electric field $E$ on the filament, the input envelope of the electric field $\varepsilon$ of the input laser beam is written as [34]:

$$\varepsilon(r,t,z=0) = E + \varepsilon_0 \exp(-r^2/w_0^2 - t^2/t_p^2 - ik_0 r^2/2f) \qquad (1)$$

where $w_0 = w_f(1 + d^2/z_f^2)^{1/2}$ is transverse beam waist at the distance d before the geometrical focus, the initial beam waist is $w_f = 3.5\ mm$, the laser pulse duration is $t_p = 26 fs$, the Rayleigh length is $z_f = \pi w_f^2 n_0/\lambda_0$, the lens focal length is $f = 50$ cm, and the curvature of the laser pulse at the distance d from the linear focus is $f = d + z_f^2/d$. Note that the electric fields at the tip of the electrode in the experiment is not linear. They diverge out for positive voltage and converge in for negative voltage. For simplicity, the E in equation (1) is approximated as the field along the axis of the electrode and all other diverging/converging fields are neglected since the maximum external electric field E (less than $10^8$ V/m) in the experiments is approximately 2 orders weaker than the laser electric field (~$10^{10}$ V/m). In consideration of the experimental results that improvement of white laser pointing stability is independence of the laser polarization, the laser polarization direction is set parallel to the electric field and scalar equation is used in the simulation. Simulation starts at *d*=30 cm before the geometrical focus where no significant ionization occurs.

Meanwhile, the NLSE for describing the laser pulse propagation can be expressed as [34]:

$$\frac{\partial \varepsilon(r,z,t)}{\partial z} = \frac{i}{2k_0}\Delta_\perp \varepsilon - i\frac{k''}{2}\frac{\partial^2 \varepsilon}{\partial t^2} - \frac{\sigma}{2}(1 + i\omega_0\tau)\rho\varepsilon - \frac{\beta_K}{2}I^{K-1}\varepsilon + i\frac{\omega_0}{c}n_2 I\varepsilon \qquad (2)$$

in which $k_0, \omega_0$ are the wave number and angular frequency of the laser beam. The pulse intensity is $I = c\epsilon_0 n_0|\ \varepsilon\ |^2/2$ and the input laser energy in this simulation is 5 mJ. Group velocity dispersion with coefficient $k'' = 2 \times 10^{-29} s^2/m$ [35, 36] is accounted for diffraction in the transverse plane. $\rho$ is laser ionized electron density and the collision cross section for inverse bremsstrahlung is σ =2×10$^{-24}$ m² [35]. $\beta^K = 1.27 \times 10^{-160}\ m^{17}/W^9$ [36] is related to multiphoton ionization coefficient, where $K$=10 is the minimum photon number for multiphoton ionization. Nonlinear coefficient $n_2$ is 3.2×10$^{-23}$ m²/W [36].

The evolution of plasma density distribution can be expressed as [36]:

$$\frac{\partial \rho}{\partial t} = \frac{\beta^K}{Kh\omega_0}(1 - \frac{\rho}{\rho_{at}})I^K \tag{3}$$

in which $h$ is the Plank constant and $\rho_{at} = 2.7 \times 10^{25}\ m^{-3}$ [3] is the neutral molecule density in air under one standard atmospheric pressure. On the right hand of Eq. (3), multiphoton ionization is taken into account. The averaged characteristic ionization energy $E_g$ for $O_2$ and $N_2$ is taken as 14.6 eV [35] in the simulation.

**Simulation on plasma decay time in the filament**

Under an external DC electric field, dominant processes involved in the wake of plasma channel induced by femtosecond laser pulse include electron-ion recombination, ion-ion recombination, impact ionization, attachments and dissociative attachments of electrons to oxygen molecules. Therefore, time evolutions of electron density $n_e$, positive ion density $n_p$, and negative ion density $n_n$ can be described by following equations [37]:

$$\begin{cases} \frac{\partial n_e}{\partial t} = \gamma n_e - \eta n_e - \beta_{ep} n_e n_p \\ \frac{\partial n_p}{\partial t} = \gamma n_e - \beta_{ep} n_e n_p - \beta_{np} n_n n_p \\ \frac{\partial n_n}{\partial t} = \eta n_e - \beta_{np} n_n n_p \end{cases} \tag{4}$$

where impact ionization coefficients $\gamma = \left(\frac{N}{N_0}\right)\frac{5.7\times10^8 \alpha^5}{1+0.3\alpha^{2.5}}$ [37]. N is air density, $N_0 = 2.688\times10^{25}$ m$^{-3}$, $\alpha = 3.34\times10^{-7}E\frac{N}{N_0}$, and $E$ is strength of external DC electric field. Attachment coefficient of electron to $O_2$, $\eta = 2.5\times10^7 s^{-1}$. The coefficients of electron-ion recombination and ion-ion recombination are $\beta_{ep} = 2.2\times10^{-13} m^3/s$ and $\beta_{np} = 2.2\times10^{-13} m^3/s$, respectively [37].

**Simulation on filament position change under DC electric field in space**

In order to calculate the electric field distribution around the electrode tip under different voltages, finite element analysis was used to solve the Poisson equation and the Maxwell equation:

$$\nabla(\varepsilon_r \varepsilon_0 \nabla\varphi) = -\rho_V \tag{5}$$

$$E = -\nabla\varphi \tag{6}$$

where $\varepsilon_r$ is the relative dielectric constant, $\varepsilon_0$ is the permittivity of vacuum, $\varphi$ is the electric scalar potential, and $\rho_V$ is the density of volume charges.

The plasma filament is assumed to contain $N$ electrons and positive ions (filament is electrically neutral)). The light electrons inside the filament tend to move toward the positive electrode, while the heavy positive ions are pushed slightly away leading to the charge separation. For simplicity, the distances from the electrons and positive ions to the electrode tip are $l$ (~1 mm in the experiment) and $l+a$, respectively, where $a$ is the diameter of plasma filament (0.1 mm) as shown in Fig.1.

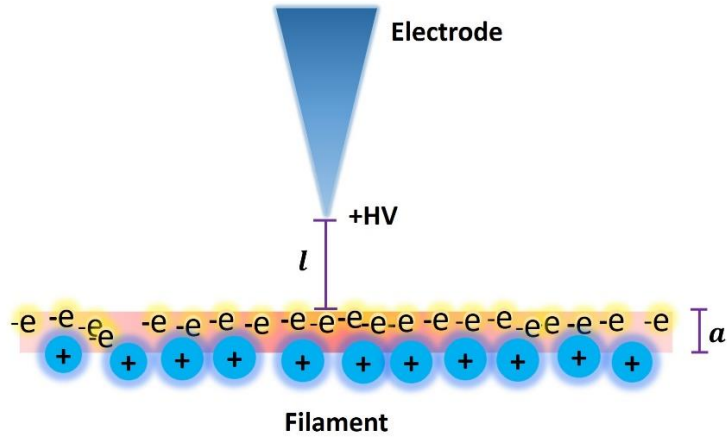

Fig.s2 Schematic diagram of the electrode applied filament

According to the Coulomb interaction equation, the force (F) imposed by the electrode tip (equivalent to point charge $Q$) on the plasma filament and the corresponding position change ($S$) of the filament are:

$$F = \frac{Q*N*e}{4\pi \varepsilon_r' \varepsilon_0 (l)^2} - \frac{Q*N*e}{4\pi \varepsilon_r' \varepsilon_0 (l+a)^2} \quad (7)$$

$$S = \frac{F}{2m} t^2 \quad (8)$$

where $m$ is the total mass of nitrogen ion and electron, $t$ is the lifetime of plasma channel under the external electric field. When the applied voltage on the electrode is higher than the threshold for coronas, the electrons inside the plasma filament would be neutralized by the corona electrode and decreases exponentially $N\sim\exp(-U)$, where $U$ is the applied voltage. Initial position of the filament without applying voltage is defined as '0'. When laser filament moves closer to the electrode, the distance value is negative.

## Acknowledgements

This work was supported in part by the Strategic Priority Research Program of the Chinese Academy of Sciences (Grant No. XDB16010400) and the International Partnership Program of Chinese Academy of Sciences (Grant No. 181231KYSB20160045). SLC acknowledges the support of COPL, Laval University, Quebec City, Canada. The authors thank Prof. Howard M. Milchberg from University of Maryland for the fruitful discussions and his reading of the manuscript.

[20] Guillaume Point, Carles Milian, Arnaud Couairon, Andre Mysyrowicz, and Aurelien Houard, "Generation of long-lived underdense channels using femtosecond filamentation in air", J. Phys. B 48, 094009 (2015)

[21] V. P. Kandidov and S. A. Shlenov, "Thermal self-action of laser beams and filamentation of pulses in turbulent atmosphere", Atmospheric and Oceanic Optics 25, 192 (2012).

[22] J. Yang, T. Zeng, L. Lin, and W. Liu, "Beam wandering of femtosecond laser filament in air", Opt. Express 23, 25628 (2015).

[23] Elise Schubert, Lorena de la Cruz, Denis Mongin, Sandro Klingebiel, Marcel Schultze, Thomas Metzger, Knut Michel, J. Kasparian, and Jean-Pierre Wolf, "Dual-scale turbulence in filamenting laser beams at high average power", Phys. Rev. A 94, 043808 (2016).

[24] N. Jhajj, Y.-H. Cheng, J.K. Wahlstrand, and H.M. Milchberg, Optical beam dynamics in a gas repetitively heated by femtosecond filaments, Opt. Express **21**, 28980 (2013)

[25] C. G. Durfee III, and H. M. Milchberg, "Light pipe for high intensity laser pulses", Phys. Rev. Lett. 71, 2409 (1993).

[26] H. M. Milchberg, T. R. Clark, C. G. Durfee III, and T. M. Antonsen, "Development and applications of a plasma waveguide for intense laser pulses", Phys. Plasmas 3, 2149 (1996).

[27] Huailiang Xu, Ya Cheng, See-Leang Chin, and Hong-Bo Sun, "Femtosecond laser ionization and fragmentation of molecules for environmental sensing", Laser & Photonics Rev. 9, No. 3, 275–293 (2015).

[28] Kai Wang, Benjamin D. Strycker, Dmitri V. Voronine, Pankaj K. Jha, Marlan O. Scully, Ronald E. Meyers, Philip Hemmer, and Alexei V. Sokolov, "Remote sub-diffraction imaging with femtosecond laser filaments", Opt. Lett. 37, 1343-1345 (2012).

[29] H. Varel, D. Ashkenasi, A. Rosenfeld, M. Wähmer, E.E.B. Campbell, "Micromachining of quartz with ultrashort laser pulses", Appl. Phys. A 65, 367–373 (1997).